\documentclass{article}

\usepackage{graphicx} % more modern
\usepackage{subfigure} 
\usepackage{amsmath} 
\usepackage{amssymb} 
\usepackage{gensymb} 
\usepackage{natbib}
\usepackage{setspace}

\begin{document} 

%%\twocolumn[
\title{Free flight of the mosquito \textit{Aedes aegypti}}
\author{S.M. Iams}
%\address{Center for Applied Mathematics,
%Cornell University,
%Ithaca, NY,
%USA.}
%\email{smi6@cornell.edu}
%
\maketitle
\section{Summary}
High speed video observations of free flying male \textit{Aedes aegypti} mosquitoes,
the dengue and yellow fever vector,
along with custom measurement methods,
enable measurement of wingbeat frequency, body position and body orientation 
of mosquitoes during flight.
We find these mosquitoes flap their wings at approximately $850$ Hz.
We also generate body yaw, body pitch and wing deviation measurements 
with standard deviations of less than $1\degree$ and
 find that sideways velocity and acceleration are important 
components of mosquito motion.  
Rapid turns involving changes in flight direction
often involve large sideways accelerations. These
do not correspond to commensurate changes in body heading,
and the insect's flight direction and body heading are
decoupled during flight.
These findings call in to question the role of yaw control in mosquito flight.
In addition, using orientation data, we find that sideways accelerations 
are well explained by roll-based rotation of the lift vector.  In contrast,
the insect's body pitch angle does not correlate with its forward acceleration.
This implies that controlling body roll is important to mosquito dynamics.
The dynamic importance of stabilizing body pitch and body yaw is less clear.

\section{Key Words}
mosquito, flight, frequency, body orientation, roll, sideways

\section{Introduction}

Mosquitoes are a human disease vector. They transmit dengue fever, 
malaria, and other diseases that impact tens of millions of people 
worldwide each year [1].%\cite{WHO}). 
Since mosquitoes approach human hosts in flight, and also mate in flight,
their flight abilities are central to their role as
disease vectors.  
Because of their prominence as human disease vectors,
many studies of mosquito flight
%\cite{Knols1996},
%\cite{Geier} 
%\cite{Dekker}, 
have focused on identifying compounds
that attract mosquitoes to humans [2,3,4].
In addition, some experimental work has focused on 
mosquito trapping [5].
%\cite{Braks}. 
%\cite{Knols} have examined 
Landing
site selection on human hosts has also been examined [6].
There is also recent work focusing on observing flight trajectories of 
individual mosquitoes [7], and on identifying flight distance and dispersal [8].
%\cite{Beeuwkes}).
%\cite{Harrington}
Very little is known, however, about the flight kinematics 
of mosquitoes. 
Making kinematic observations of flying mosquitoes has the potential
to inform our understanding of the insect's sensory
experience during flight.  Such observations also make it possible 
to examine stability and control during flight.

There are few observations of insect orientation during free flight.
Because of this,
the extent to which insects manipulate their body pitch to generate forward thrust
or their body roll to generate sideways thrust is not known.  
The relationship between body heading and flight heading is also
not well characterized.
In addition to these general questions, the geometry of the mosquito 
during flight is distinct from that
of commonly studied small dipterans, so it is unclear to what extent observations
of hoverflies or fruit flies translate to the mosquito.  
Mosquito bodies are long and slender compared to fruit flies, 
and they have long legs, as long as their
thorax and abdomen combined, that are held outstretched during flight.
They have an unusually short and quick wingstroke, 
subtending just $45 \degree$ with their wings on each half stroke, compared with $120 \degree$
for fruitflies.  
This significant kinematic difference, combined with a very different body geometry,
 means that, although mosquitoes likely share flight
control strategies with other dipterans, 
they potentially utilize different strategies as well.

To aid in identifying the mosquito's flight capabilities, we have developed 
methods to make automated measurements of body position and orientation.
With these position and orientation measurements we examine
how mosquitoes orient themselves during steady flight, as well as during maneuvers.
We measure the heading of the insect as it flies.  Since the
orientation of the head dictates the direction of visual, auditory, and odor sampling
via the eyes and antennae, it is interconnected with their sensory experience and is particularly
important to understand.
In addition, we investigate the relationships between pitch, roll, and body acceleration during
flight.

For small insects, measuring body orientation during flight requires processing flight images
to extract proxies for orientation.
For larger insects, such as the blowfly, direct measurements via instrumentation attached to
the insect are possible.
Flight orientation,
including body yaw, pitch and roll, has been carefully documented during nearly free
flight for blowflies 
%\cite{Schilstra} 
by attaching a pickup coil to the insects as they fly
in a magnetic field [9].
The blowflies they studied weighed $80-100$ mg, about $80$ 
times as much as small male \textit{Aedes aegypti},
and had a wingbeat frequency of approximately $140$ Hz, 
compared to approximately $800$ Hz for an \textit{Aedes aegypti}
male.
Blowfly flight is distinguished by periodic large changes in body orientation, termed saccades
because these body motions correspond to rapid motion across the visual field.  
In their measurements, most saccades lasted less than $30$ ms, and occurred at regular intervals,
happening approximately
every $100$ ms. The body heading of the insect changed slowly outside of these times.
In addition, during turns, the blowfly moved its head relative to the thorax, so that head saccades occurred
even more rapidly than thoracic ones
[10].
%(\cite{Hateren}).  
Observed saccades
ranged from $10-90$ degrees, with smaller changes being more common, and occurred in the context of
banked turns, where the insect's body orientation and its flight heading change simultaneously.
These turns involved changes in yaw, 
and blowflies also vary their pitch and roll angles as part of actuating saccadic banked turns.
The blowfly flight repetoire extends beyond saccadic turns.   
They are able to make U-turns, where large roll angles produce sideways accelerations,
and have also been observed flying backwards into a turn. 
This blowfly data is unique in including direct measurements of body roll.

Hoverflies 
are a small dipteran, only slightly larger than small male mosquitoes. 
Their body heading during flight, and during a number of 
flight maneuvers, has been well documented 
[11].
%\cite{Collett}.  
Hoverflies have been observed flying 
forward, sideways, and backward, as well as   
abruptly changing the orientation of their body axis via saccadic turns
without changing their flight direction.
Thus their body axis and flight direction are not
necessarily aligned during flight.  
Although not always aligned, the two often are
[12].
%(\cite{Geurten}).
In addition to rapid saccades, hoverflies are also capable of continuously 
changing their body orientation in so-called smooth tracking, where an object of interest is aligned in the center
of their visual system instead of at the periphery.  Such smooth changes in body orientation can be induced by
showing the insect a rotating pattern of stripes.
In free flight, hoverflies have a repertoire of maneuvers that includes banked turns, resembling those of blowflies,
although body roll and pitch information are not available for hoverflies during such turns.
 Hoverflies also turn by adding
a strong sideways component to their velocity, without
any change in body orientation.  We observe many of these
sideslip based turns in mosquitoes.

Fruit flies are another commonly studied small dipteran.
However, free flight measurements that include body orientation are rare.
Such measurements have been made during yaw turns
[13, 14]. 
%\cite{Bergou},
%\cite{Fry}.
Free flight data for fruitflies is often recorded at a coarser scale, looking at
 flight path trajectories without measuring body orientation [15, 16, 17, 18, 19].
% \cite{Tammero}, \cite{Mronz},
%\cite{Budick} and \cite{Reynolds}.
%\cite{Beurgen}
In studies of flight trajectories in fruit flies, flight direction 
is sometimes used as a proxy for body orientation.  
In the context of analyzing flight paths,
maneuvers where the insect rapidly changes its flight direction
have been classified as body saccades.  
It is unclear whether these very fast changes in direction
are accompanied by similarly fast changes in body orientation in fruit flies.

We take time-resolved free flight measurements of \textit{Aedes aegypti} males at 
sufficient spatial resolution so as to observe body orientation and wing motion.  Over the course 
of $4000$ wingbeats of flight data, we observe a number of turning maneuvers,
where the flight direction of the mosquito changes substantially.  
These maneuvers are not experimentally induced, but are freely initiated by the insects.
We observe sideslip based turns that involve abrupt changes in flight direction, turns that
involve large changes in body orientation, and 
gradual U-turns, similar to those in blowflies, 
where flight direction changes by approximately $180$ degrees.  
We find that these mosquitoes
most frequently fly with their body orientation and flight path direction unaligned, 
meaning sideways acceleration is an 
important part of their flight repertoire.
In addition, by examining $100 ms$ flight segments,
we observe many changes in body
orientation that are more gradual, along with a few abrupt changes that may be considered saccades, 
occurring both during and outside of turns. 
Finally, we examine pitch and roll information and find that roll angle correlates
strongly with observed sideways acceleration, but there is no consistent correlation
between body pitch and forward acceleration.

\section{Materials and Methods}

\begin{figure}
{\includegraphics{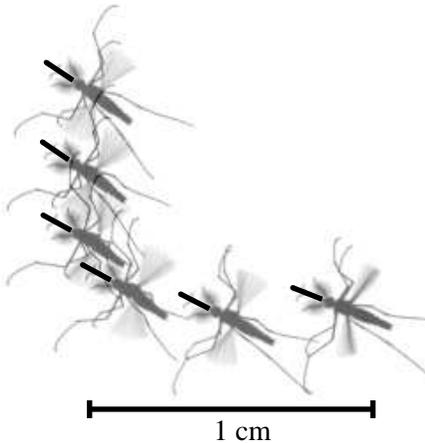}}
\caption{
To illustrate a turning maneuver, 
snapshots of a mosquito are shown at $30$ ms intervals, corresponding
to approximately $25$ wingbeats. 
The wings are blurred because each image shown is the sum of five consecutive frames.
Although the turn involves a rapid change in flight direction, the insect's heading
evolves much more slowly, and does not change very much.  
Measurements of proboscis position and orientation
are represented by a black line superimposed over each mosquito image.
}
\label{tracking2}
\end{figure}

\subsection{Flight observation}
Our observations of mosquito flight involved recording free flight in four to seven day old unmated male 
mosquitoes. These males weighed between 0.75 mg and 1.1 mg, with a typical body length of 
approximately $5$ mm. Using a mouth aspirator, a number of mosquitoes 
were transferred to a plexiglass flight chamber.
The chamber measured $15$ cm on each side, and was placed in the center of three perpendicularly 
aligned high speed cameras. The flight recording
setup was as in [20],
%\cite{Ristroph}, 
with one overhead and two side views, and insect motion was
captured as shadows against
a lit background.
The cameras were synchronized and were triggered by hand.

Male \textit{Aedes aegypti} fly with a wingbeat frequency of approximately $800\ Hz$. 
We recorded these insects at
$13000$ frames per second, so that each wingbeat could be resolved. 
Although the flight chamber was large, 
 due to limitations of resolution at such high speeds,
we only imaged a $2.4\ cm$ cube located near the center of the flight arena. Thus,
many flights within the chamber were not recorded. Mosquitoes are active fliers, 
however, so it was not rare for them to pass
through the central recording region.

In some recorded flights, an earbud speaker was located just outside of the recording area. Pure 
$400\ Hz$ tones were played as a flight stimulus. The female \textit{Aedes aegypti} produces a 
fundamental flight tone near this pitch,
 and such pure tones act as a stimulus for male mosquito flight [21].
%\cite{Cator}.
In other flights, recorded without a sound stimulus, a brief exposure to human breath primed the mosquitoes to initiate flight.
In addition, flight was sometimes stimulated via a looming stimulus, with an object from outside the flight chamber
approaching a stationary mosquito.

With many mosquitoes in the flight chamber at the same time, we were unable to identify individuals. Thus it is unknown how
many distinct individuals we observed in flight. We recorded over $40$ sequences of free flight, ranging in length
from just a few wingbeats to over $600$ wingbeats. 
These represent more than $4000$ wingbeats of flight time, about $5 s$.

\subsection{Measurement Methods}
\begin{figure}
{\includegraphics{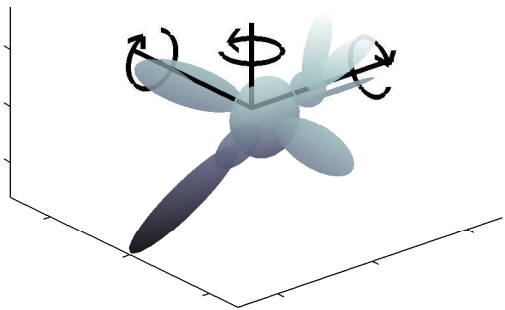}}
\caption{
In a coordinate frame attached to the body, the body x-axis aligns with the head and
proboscis.  When the insect is neither rolled nor pitched, the body z-axis is
nearly vertical, and the body y-axis is perpendicular to these, lying along the wing stroke plane.
In this frame, we use the xyz convention to describe body orientation. 
We call rotation about the $z$-axis 'yaw', rotation about the $y$-axis
'pitch', and rotation about the $x$-axis 'roll'.
To transform from a world frame sharing its origin with the body frame,
we first roll about the world fixed $x$-axis, then pitch about the fixed $y$-axis
and finally yaw about the fixed $z$-axis.  In axes moving with an object, the object
would first yaw about its local $z$-axis, 
then would pitch about its local $y$-axis, and finally would roll about its local $x$-axis.  
These transformations are reversed
to transform from the body frame back to the world frame.
}
\label{figcoords}
\end{figure}

We think of the insect as a rigid body with three spatial and three
rotational degrees of freedom.
We define our axes using the xyz convention, also referred to as yaw, pitch and 
roll, as in [22].
%\cite{Goldstein}.  
From the lab frame,
using a coordinate frame with its origin centered between the two wings,
we assign the center line of the mosquito to be the body x-axis.  This passes through
the thorax and proboscis, and is the axis about which the insect rolls.
With the mosquito in a neutral position, so that its stroke plane is neither pitched nor rolled, 
the z-axis is vertical.  This is the axis about which the insect yaws.  
With a right handed coordinate system, the y-axis is perpendicular to the xz-plane and points in the direction of the left wing.
The insect pitches about this axis.
 These definitions are shown in figure \ref{figcoords}.
To gain insight into the insect's motion, we measure the yaw (body heading),
pitch, and roll.  We also estimate the position of the body in each frame of the movie.

Orientation measurements are important to reconstructing the sensory environment
of the insect, and for gaining insights into insect flight dynamics.
Position measurements enable us to estimate velocity, acceleration, and other
statistics about the flight path, and are important to understanding
the aerodynamic environment the insect experiences.
Since the eyes and antennae are attached to the heading, knowing the head
orientation informs our understanding of the visual and sensory environment.
The wings attach to the thorax, so changes in thoracic orientation alter
the orientation of the stroke plane of the insect and this the direction
of flight forces.  We assume that the head, thorax and abdomen do not
move in relation to each other over the short periods of flight that
we record; thus we use orientation measurements based on the proboscis
and the abdomen as estimates of thoracic orientation.

We use feature-based methods, supplemented with reconstruction-based measurements,
to gain information about the mosquito's position and orientation.
To analyze each flight sequence, we 
start by subtracting each image from a bright background image in which the 
shadow of the mosquito is missing, leaving an image of a bright mosquito
on a black field.  This process creates the frames we will use for analysis.
In our feature-based measurement methods, we measure the location of a feature on the insect, 
confirm our measurement and then use this validated information to initialize our next measurement.
To start this process, we create information streams about the images from which we
can extract phase and frequency information, and we track the locations of the proboscis tip and
abdomen tip in the overhead view.  
Once validated, we use these initial phase measurements to identify appropriate frames
for further analysis, and we use the position measurements to initialize other feature
trackers.

More detailed descriptions of some of the measurement methods, as well as the details
of our measurement error analysis, are included in the supplementary materials.

\subsubsection{Identifying Phase}
During each wingbeat, there are times
when the wings are translating rapidly, and times when they are nearly stationary.
In addition, the wings rotate during flight, so that
the wing sometimes has a narrow profile and sometimes a wide profile.
During times of rapid wing translation, if we substract the previous frame from the current
frame, we isolate the leading portion of the wing, and can similarly isolate the trailing portion
using the next frame.  
We may capture non-wing pixels if the body is translating at particularly high speed
or a leg is in motion, however, the wing pixels form the major component of differenced
images.  
  At times of rapid rotation
but low translation, these frame differences isolate very few pixels as the wing profile
is narrowing, and capture more as it is widening.  
Counting the number of bright pixels in the
differenced frames generates a signal that reflects the periodicity of the wing motion.
This periodic signal might be superimposed on a background signal that reflects leg
or body motion, however the periodic portion of the signal is due to wing motion.

Given this periodic signal, we use a published phase identification algorithm 
%\cite{Revzen} 
to estimate the instantaneous phase
of the wingbeat [23].  
With this phase information, we are able to calculate flapping frequency based
on just a few wingbeats, so that it is possible to observe the evolution of the
frequency over time.
In addition, this phase information depends on both wing translation and rotation,
and encodes where the wings are in their stroke.  

\subsubsection{Feature Tracking}
We do all initial tracking in the overhead view.  Although the phase identification
methods were fully automated, these feature trackers are hand initialized.
In the first frame of a flight sequence,
we hand identify the tip of the proboscis
 and the tip of the abdomen.
We refine these hand-identified locations using a cross-correlation method.
These refined measurements act as the initial measurement in the next frame.
the refinement procedure is carried out again, and we
continue to step forward using this simple procedure.
This straightforward tracker is error prone only when an occlusion occurs,
such as a leg crossing the abdomen or proboscis.  These are rare events,
are easily identified by watching a flight video with the tracked points
superimposed, and can often 
be corrected in an automated way by tracking frames backwards instead of forwards.  
We find that this method identifies a point along the centerline of a simulated proboscis with
a standard deviation of less than $0.1$ pixels.  
The vector between the proboscis tip and the abdomen tip is then used to initialize our
other tracking methods.

\begin{figure}
{\includegraphics{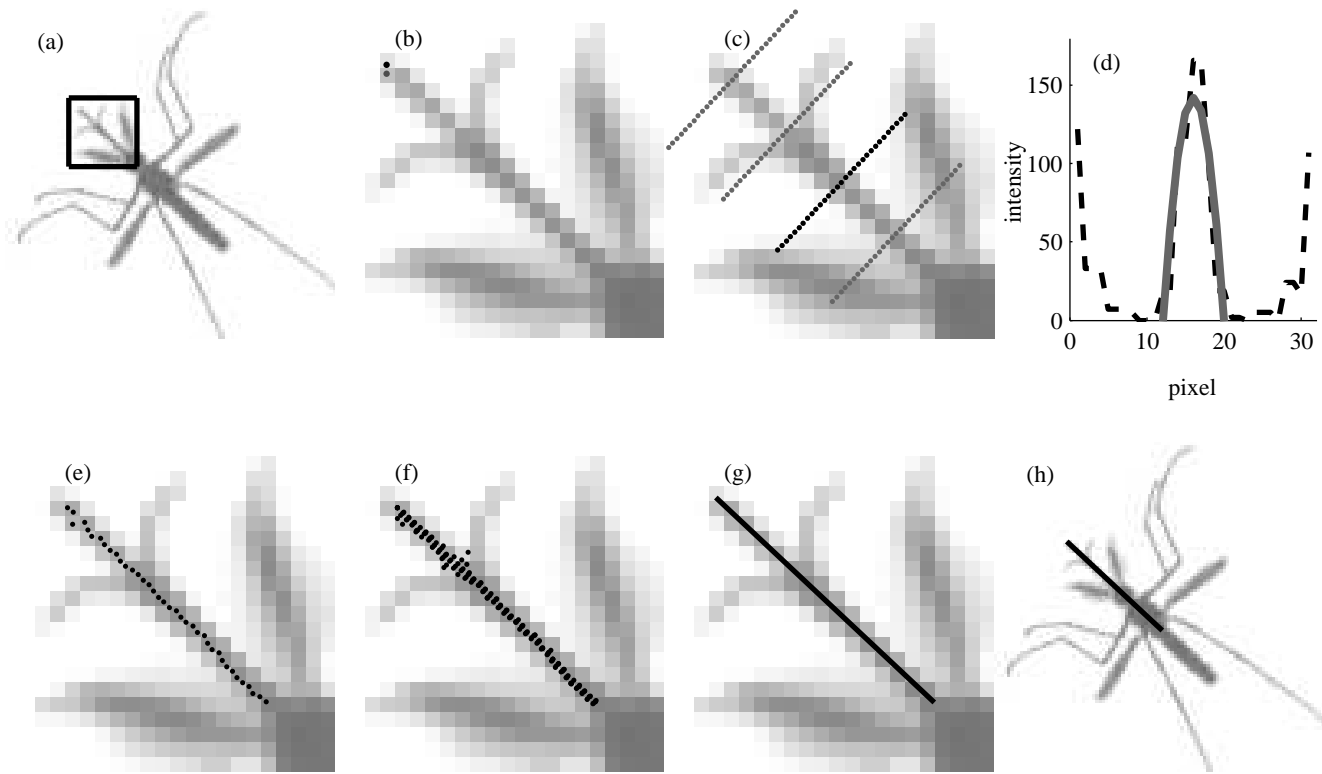}}
\caption{ 
The proboscis measurement method is shown in detail here.
In (a), we show the region of the insect that we are focusing on during these
measurements.  In (b), the preliminary location of the proboscis tip is shown in
dark gray.  It is slightly below the refined position, which is shown in black.
The refinement procedure uses cross correlation and is described in the supplementary
materials.  Walking along the vector direction between the proboscis tip and the abdomen tip,
we are able to take cross sections intersecting the proboscis.  Four of these
cross sections are plotted in (c).  The cross section in black corresponds
to the intensity profile (in black) in (d).  In gray in (d), we
plot our quadratic fit to the center of this intensity profile.  The calculated
peaks of the quadratic fits to many cross-sections are plotted as points in (e).
We show five successive frames of this data pooled together in (f).  Using all of these points,
 we calculate a line of fit to the measured points.  This line is shown in
(g).  In (h) we shown the alignment of this line on the original image.
The orientation of this line of fit corresponds to the body heading of the insect.
}
\label{heading}
\end{figure}

To measure the body heading, we measure the orientation of the proboscis in the overhead view.
Although the vector between the proboscis tip and the abdomen tip may appear to align with the
body heading,
mosquito body geometry is such that
a non-zero thoracic roll
results in the sideways deflection of the abdomen.
When this deflection occurs, the tip of the abdomen no longer lies along the central axis of the insect
as seen from the overhead view.  
We use the orientation of the proboscis
as a less error prone estimate of the insect’s heading.
The measurement procedure is described in figure \ref{heading} and is discussed further in
the supplementary materials.
Our heading measurements have a standard deviation of less than $0.2$ degrees
when compared to known proboscis orientations in synthetic images.

To estimate the insect's roll angle, 
we use information from our phase measurement, as well as the body heading.
Very little quantitative data exists on roll during free flight, because roll is difficult to
measure from features of the body.  We estimate roll by first measuring the position of each wing vein
at the center of the midstroke and then calculating the deviation of each wing from horizontal. 
We average these deviations to generate our roll estimate. 
The appropriate midstroke frames for these measurements are identified by their phase.
We look for the frames where the wing veins come close to forming a line perpendicular 
to the body heading.  This occurs once on the forward stroke and once on the backstroke.

To identify points on the wing vein, we first find the leading edge of the wing in the overhead
view, and identify points along that edge.  
Once they are identified, we find corresponding points in the xz or yz view to locate the edge
of the wing vein in three space.
We project this vein into the yz plane of the body, and measure the deviation angle of the vein
from horizontal.
This measurement process is described in more detail in the supplementary materials.

When the
insect is rolled, one wing tilts upwards from the horizontal
and the other downwards.  
Positive roll corresponds to the left wing moving upwards
and the right wing down.  We average the left wing angular deviation measurement
with the negative of the right wing measurement to estimate roll.
Our measurement method identifies the angular deviation of a straight edged wing
from horizontal with a standard deviation of less than one degree.

\subsubsection{Body Reconstruction}

We use reconstructive methods to determine the body pitch and the location of the
body centroid.  These are similar to methods described in [20].  This involves
creating a voxel reconstruction of the body or the abdomen.  We find the centroid
of the body voxels to determine the body location.  To measure the pitch,
we find the first principal component vector along the abdomen voxels and
measure its inclination away from horizontal.  Further details of the algorithm are contained
in the supplementary materials.  The body pitch methods results in
measurements that are systematically slightly high and have a standard deviation
of less than $0.5$ degrees.  The body centroid measurements identify a consistent
point on the body within one pixel.

\subsection{Finding derivatives}

The relationship of the insect's body heading to its flight direction,
where the flight direction is defined as the direction instantaneously
tangent to the flight path, is important for reconstruction of the visual
stimuli an insect experiences.  It is also important to understanding
flight dynamics, because offsets between the body heading and the flight
direction alter the direction of the oncoming flow of air.  This means
that the two wings may not be in a symmetric flow field, even
when moving symmetrically.  To measure this offset angle, we must
identify the direction of the flight path.  This direction corresponds
to the vector direction of the insect's instantaneous velocity.
To determine the velocity of the insect, we use the functional
data analysis package for MATLAB discussed in 
[24].
%\cite{Ramsay}.  
Using cubic splines, so that the insect’s acceleration is piecewise linear
over each wingbeat, and setting a condition so that the curvature of the 
second derivative is low while the splines fit the data well,
we create a functional fit to the insect’s body position.
We can then estimate the body velocity by taking the first derivative of this 
fitted function.  It is similarly possible to estimate
the body acceleration by taking the second derivative.  
To determine the direction of the flight path,
we are able to find an angle corresponding to the vector
direction of the body velocity in the horizontal plane.

\section{Results}

\begin{figure}
{\includegraphics{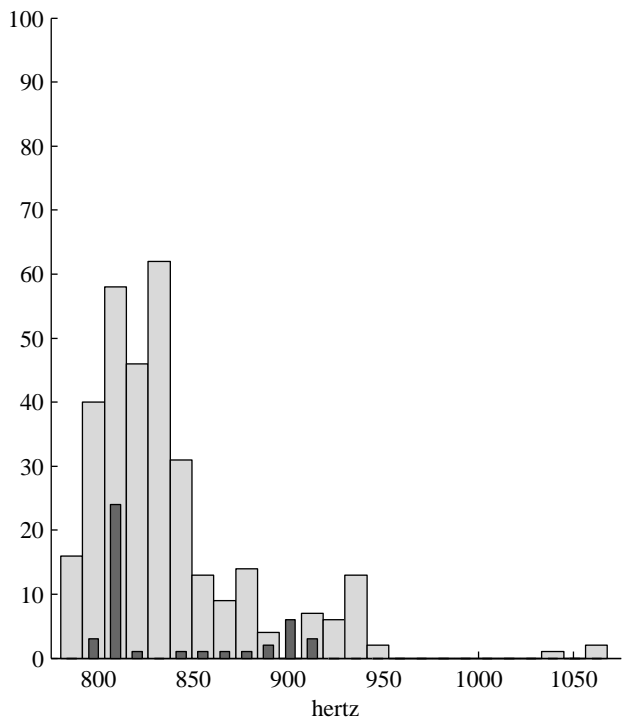}}
\caption{
Frequency during flight was averaged over $10$ wingbeat periods to create this histogram.
Dark gray bars indicate flapping frequencies during flights in the presence of a $400$ Hz
sound stimulus, while in light gray are flight frequencies for flights without such a stimulus.
In free flight, male \textit{Aedes aegypti} typically flap at approximately $800$ Hz.
}
\label{freq}
\end{figure}

\begin{figure}
{\includegraphics{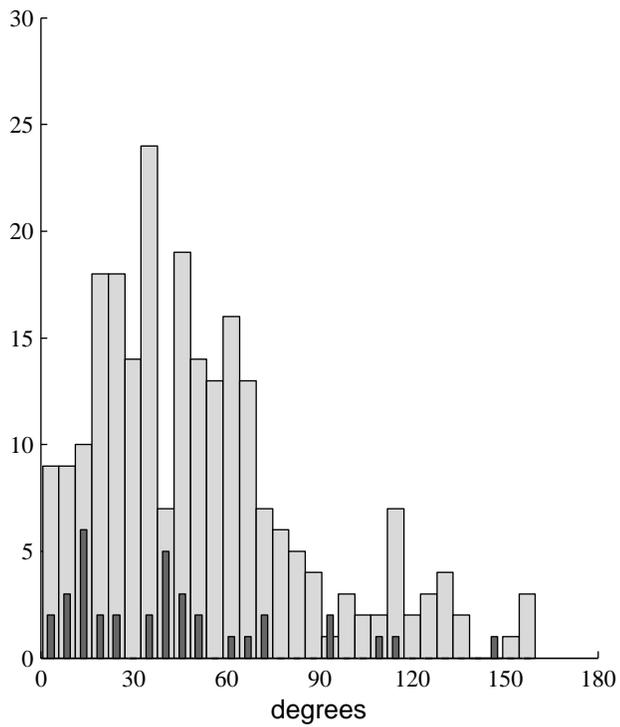}}
\caption{
These angles represent the difference, measured in degrees, between the insect's
body heading and its instantaneous flight direction.
This angle is sampled every $20$ ms, which is equivalent to approximately $17$ wingbeats.
Dark gray bars indicate represent data from flights that occurred in the presence of the
sound stimulus, while light gray bars come from flights without such a stimulus.
The distribution is peaked away from $0\degree,$ indicating that most flight
have a sideways component, and has long tails that represent flights that
are predominantly backwards.
}
\label{orient}
\end{figure}

\begin{figure}
{\includegraphics{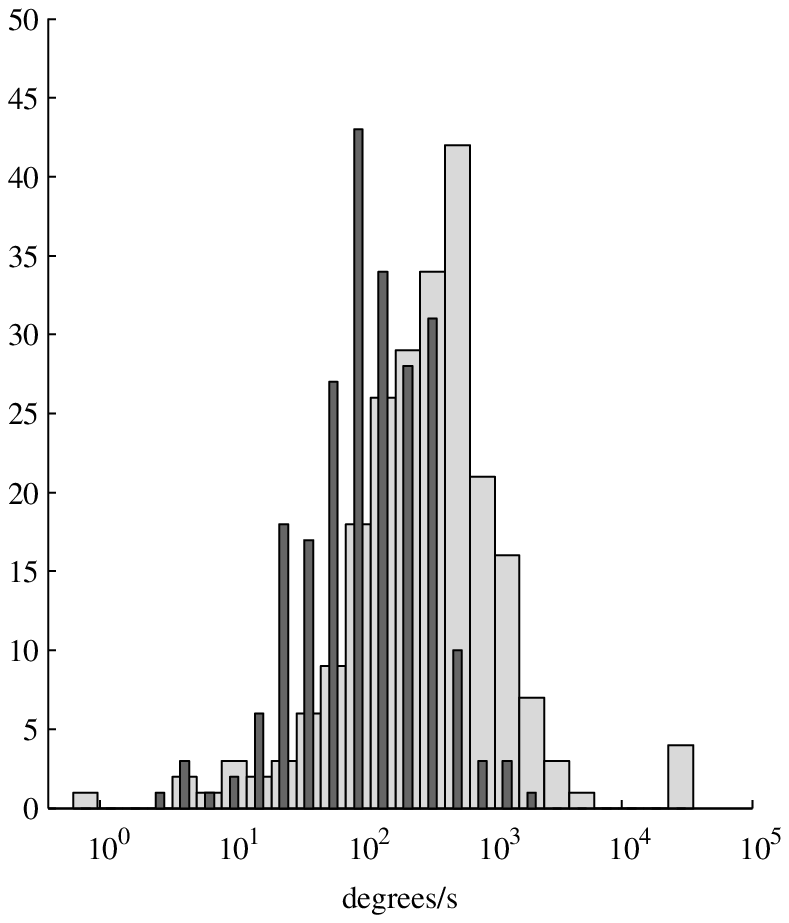}}
\caption{
These histograms show
the rate of change in body heading angle in dark gray, and
the rate of change in flight direction in light gray.  
These angular velocities are sampled every $20$ ms.  Both the flight
direction and the body heading are typically changing by more than $10\degree$
on the timescale of a second.  However, changes in body orientation are
almost an order of magnitude less rapid than changes in flight direction.  
}
\label{angular}
\end{figure}

In the presence of a 400 Hz sound stimulus, 
we observed the mosquitoes flying in nearly straight track segments.  Since
the flight chamber was limited in size, the mosquitoes did not achieve flight velocities above $0.4 \frac{m}{s}$
during these flights.
Their wingbeat frequencies, calculated as average frequency over ten wingbeat flight segments, 
exhibited a large spike at just above $800$ Hz and a smaller spike near $900$ Hz.
The larger spike occurs at approximately double the ambient $400$ Hz stimulus.
The distribution of frequencies
during sound stimulated flight is shown in dark gray in the histogram in figure
\ref{freq}.  For flight without background sound, shown in light gray, flight frequency was much more
broadly distributed.  Frequencies ranged
from 780 Hz to 950 Hz, with rare observations of frequencies as high as 1050 Hz.  Within a single flight,
we observed frequency changes of up to 50 Hz.

During sound stimulated flight, the mosquitoes typically flew in straight lines.
During these flights, though, their body heading rarely aligned exactly with their 
flight direction.
Instead,
as shown in dark gray in the histogram in figure \ref{orient},
flights ranged from nearly forward, to fully sideways, and to almost backwards.
This wide range of flight directions also occurred during flight
without background sound.  This 
is shown in light gray in the same histogram.  
For this mosquito dataset, the distribution of body heading relative to flight direction 
is somewhat flat from $0\degree$ to $75\degree,$
with no clearly preferred direction.
In hoverflies, a similar
histogram is more sharply peaked about zero [11].  
%\cite{Collett} 
This is
also the case for honey bees [25].
%\cite{Boeddeker}.  

During free flight without sound, we observed a number of turning maneuvers.
These are sometimes accompanied by gradual changes in body heading
and at other times by abrupt changes.
Observed changes in body heading did not exceed 30 degrees in 50 ms.
At the same time, body heading was rarely stable for more than short
periods of time.  As shown by the dark gray bars on the semilog axis
in figure \ref{angular}, typical rates of change for the body heading
range from 30 degrees per second
to 400 degrees per second, and are infrequently outside of that range.
%\cite{Collett} 
Saccadic turns, which might occur when the insect is tracking
something that is rotating too rapidly to be smoothly tracked, so that
rapid readjustments of the visual direction are required, typically occur
in approximately $40-460$ ms, and might have a magnitude ranging from $10\degree$
to over $100\degree$ [11].  We observed only a few such saccadic turns, each
of which was small in amplitude.

Although changes in body heading were typically gradual,
we observed many rapid changes in flight orientation.  The 
rate of change of the flight direction is shown by the light
gray bars in figure \ref{angular}.  These changes are frequently
an order of magnitude larger than changes in body heading.
Tammero and Dickinson %\cite{Tammero}
 use a criterion of a change in flight direction of 
approximately $90\degree$ in $100$ ms, which
they assume is accompanied by an identical change in body heading, 
to define a saccade [15].  Mronz and Lehmann,
%In \cite{Mronz}, 
making a similar assumption about 
flight direction and body heading,
used a threshold of $1000 \degree/s$ in the angular velocity of the flight
direction angle to define a saccade [16].  
It is misleading to associate saccadic turns with these criteria,
as they relate to flight direction, and not to body heading.
We observe $12$ turning manuevers consistent with these criteria.  
During these turning maneuvers, the insects change
flight heading by $50\degree$ to $200\degree$.
The smallest of these changes are similar to our largest observed changes in
body heading.  This means that when the insect changes flight direction
there is not a commensurate change in body heading, so heading and flight
direction are not well coupled.
This is similar to the hoverfly, where sideslip-based turns are common for changing
direction.
In both insects, the heading sometimes stays steady even as 
the flight direction changes dramatically.
However, the hoverfly appears to use also saccadic turns to turn its body heading
to align with its flight direction, and based on the flight orientation measurements
we present above, mosquitoes have no such alignment preference, potentially reducing
their reliance on saccadic turns.

\begin{figure}
{\includegraphics{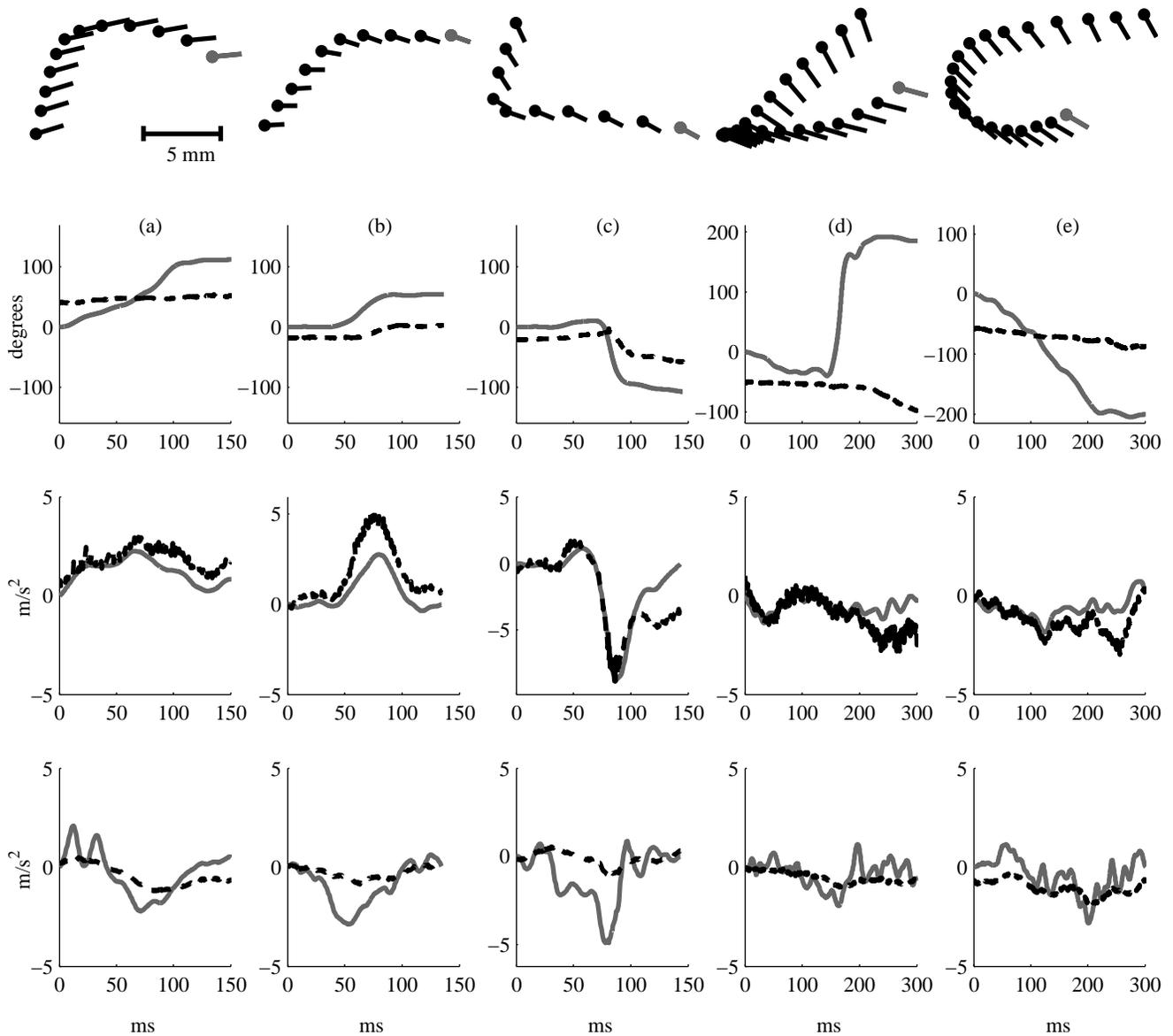}}
\caption{
In the top row, five different turning maneuvers are represented using ball and stick diagrams.
The ball corresponds to the tip of the insect's proboscis, while the stick represents its body
and is aligned with the measured body heading.  The starting point of each trajectory is
in light gray.  The insect's position and body orientation is represented every $15$ ms.
In the second row, the flight direction of the insect is shown in gray for each trajectory.
The body heading during the flight is in black.  In (b) and (c), the insect executes
saccadic turns during the course of the maneuver.  However, in both cases its flight direction changes
much more dramatically than its body orientation.
In the third row, the sideways
component of the insect's measured acceleration (in gray) is compared to the acceleration generated by a simple
model of thrust, where the lift vector is redirected by the insect's body roll to generate
a sideways thrust component (in black).  There is a strong correlation
between the roll-generated thrust and the measured sideways thrust.  In the fourth row, 
a similar simple model, 
with pitch taking the place of roll (in black), is compared to measured forward accelerations (in gray).
We found no consistent correlation between pitch generated thrust and measured forward thrust.
}
\label{turns}
\end{figure}

In figure \ref{turns}, we present five examples of turning maneuvers, with
each column of data corresponding to a separate maneuver.  In the first row,
the stick and dot diagrams depict the heading of the insect body, with the dot
at the tip of the proboscis, as it flies through the manuever.  The starting position
at the beginning of the flight segment
is indicated in gray.  In the second row of the figure, the change in flight direction
over the course of the maneuver is shown in gray.  Body heading during
the maneuver is depicted by a dotted
black line.  These lines have been recentered so that the initial flight angle of the insect
is set to zero.
Maneuver (a) is an example of a time when the insect
changes its flight direction significantly, but its change in heading is a factor of 10 smaller.
Although sometimes heading changes are very small and happen slowly, at other times they are 
larger and occur abruptly.  In 
maneuvers (b) and (c), the body heading changes significantly during the flight.  
However, the magnitude of the change is still only 1/2 to 1/3 that of the change in flight direction.  

In every maneuver observed, the insect enters the maneuver flying with a sideways velocity
component.  This means that its gaze is centered either to the right or to the left of
its flight path.  It is often the case that during the maneuver the insect moves into the
region its head is facing.  However, this 
is not universally the case and a counter-example is shown in (b).  
During most observed U turns, two of which are shown in (d) and (e), the insect flies
foward during the turn, thus moving into the region where its gaze was directed.  However,
even in U turns the insect sometimes flies backwards instead of forwards during the center portion
of the turn.
Similar U turns are seen in blowflies [9], 
%\cite{Schilstra}
and backwards motion has been observed in fruitflies [26].
%\cite{David}

To actuate turns, an insect must change the direction of its velocity, and
may change the magnitude as well.  We decompose the insect's acceleration in the horizontal
plane into a forward component that is aligned with the body heading and a sideways
component perpendicular to the heading.  We can think of the insect as producing
some combination of forward and sideways acceleration to generate the turn.
For each maneuver, the measured sideways acceleration of the body is plotted in
gray in the third row.  The measured forward acceleration is in gray in the fourth row.
During these turns, the insect applies some braking along its forward direction.
This appears as a negative acceleration in row 4.  The presence of braking means
the change in direction occurs more abruptly, and the degree to which the insect brakes
determines how much of a forward velocity component is present, so it helps set
the angle of the turn.  During these rapid turns, sideways acceleration is particularly 
important. In the higher speed, sharper turns shown in (a), (b), and (c), we see sideways 
accelerations of 0.3g to g.  Accelerations remain lower during the 
U-turns in (d) and (e) where the insect is moving at a lower velocity through the maneuver.

We estimate the pitch and roll of the stroke plane so that we can compare the measured forward
and sideways accelerations
to those generated by a simple force model.  
In this simple model, we think of the lift vector as being deflected from the vertical 
by pitch and by roll.  Since the insect continues to generate acceleration to counter
gravity, such deflections would generate acceleration along the forward direction of $g \tan\theta$
where $\theta$ is the pitch angle and acceleration along the sideways direction of $-g \tan\psi$
where $\psi$ is the roll angle.  The negative sign appears because of the way these coordinate
systems are defined.  The accelerations that would be generated under this simple model are shown
in black in rows 3 and 4. 
In the case of roll, this model consistently
overestimates measured sideways accelerations, particularly at large roll angles.
Correlation coefficients range from
0.75 to 0.95, between acceleration due to deflection of the stroke plane via roll, and sideways acceleration.
This correlation is quite strong, and suggests that although this model is missing an important
damping component, rotation of the lift vector via roll is the dominant driver of sideways motion in mosquitoes.
A similar sideways mechanism has been observed in house flies [27] and has been posited for moths [28] 
%\cite{Wagner1}
%\cite{Zanen}
Since sideways motion plays a prominent part in mosquito maneuvering, roll is a particularly important
degree of freedom for these insects.
In constrast, for pitch, measured correlation coefficients between measured and predicted
forward accelerations do not have a consistent sign, and there is no noticable relationship.
This is in line with observations of other Diptera.  Although experiments with 
tethered insects [29] and insects in a wind tunnel [26] 
%\cite{Vogel}
%\cite{David}
have shown a relationship between the body pitch or the body angle and the forward speed,
this relationship has been found not to persist in free flight for a number of Dipterans [31].
%\cite{Ennos}

\section{Discussion}

We described the methods we use to make high accuracy measurements of the heading and
of the wing stroke plane deviation.  We also present our measurement methods for body position
and pitch angle, and explain the approximations we chose in using wing stroke plane deviation
as a proxy for stroke plane roll.  When we apply these measurement methods to our data set,
we find indications that mosquitoes usually fly so that their body heading and flight direction
are not aligned.  In this dataset, they are flying at a wide range of equally preferred 
offset angles.  
Non-forward flight has been observed or inferred in other Dipterans,
such as blowflies [9], fruitflies [26], and hoverflies [11].
% blowfly, fruit-fly, hoverfly
Thus, for mosquitoes, and perhaps for other Dipterans, it is not clear when during 
the course of a flight the 
insect's orientation can be inferred from 
its flight path alone, as is sometimes
assumed for the sake of analysis of visual inputs [16, 19].
%\cite{Mronz}, \cite{Breugel}

Mosquito flight direction changes more frequently and much more quickly than does body heading.
Although all sharp changes in heading we observed occurred during turning maneuvers,
many turning maneuvers were not accompanied by rapid changes in heading.  These turning maneuvers,
in which the flight direction of the insect changes by 50 to 200 degrees, 
involve a combination of deceleration along the direction aligned with the body axis
and acceleration in a sideways direction.  We find that the observed sideways accelerations
strongly correlate with those generated by a simple roll-based acceleration model.  However,
accelerations along the body axis do not correlate with those from a similar pitch-based model.
   
Body alignment during flight has implications for the sensory experience of the insect and also
impacts our understanding of flight control.
The insect might purposefully maintain a steady body heading during a turn.
During swarming, 
\textit{Aedes aegypti} have been described as 
flying with a characteristic figure eight to-and-fro motion [31]
%\cite{Hartberg} 
and other mosquitoes are known to maintain an
upwind heading during swarming [32].
%(\cite{Downes}).
By maintaining a steady body heading even during a turn,
the eyes, antennae, and maxillary palps remain oriented in a consistent direction.
This may be desirable when searching for the source of an odor or sound signal.
Male \textit{Aedes aegypti} have been caught in odor-based traps [33], and are known
to follow sounds signals 
Male \textit{Aedes aegypti} are known to follow sound signals
when finding a flying female with which to mate [21].
In addition, they have been caught in odor based traps and landing on humans, and are
potentially sensitive to odor signals [33, 31].
%(\cite{Cator}).
%\cite{Krockel}
%\cite{Hartberg}
In the context of search,
moths are known to maintain body heading upwind, even during turns and circuitous flights [28].
%(\cite{Zanen}).
Female \textit{Aedes aegypti} exhibit similarly circuitous flight tracks interspersed with upwind
flight during exposure to intermittent odor plumes,
%(\cite{Dekker}), 
as do \textit{Drosophila}
[4, 34],
%(\cite{Zanen2}),
although their orientations during such searching flights have not been identified.
Potentially, the orientation of the insect relative to its flight path
may depend on the type of sensory stimuli used to induce flight and the sensory
environment during flight.  Identifying times when the insect is likely to have a particular 
orientation relative to its flight path, or even measurable
modes of flight in mosquitoes, if they exist, would require further work.

There may be other reasons for maintaining a constant heading through the course
of a turn.  It is possible that the heading is not purposefully maintained,
and is even potentially not actively controlled.
It is likely that for a rapid turn, it is energetically favorable to actuate a sideslip
based turn, which uses the deflection of the lift vector, rather than a yaw
based one.   Yaw based turns use torques that are the result of asymmetries between
the forces generated by the two wings,  
while sideslip based turns use the sum of the forces from the two wings, rather than
the difference, to generate the turn.
When hoverflies use these types of sideslip turns, they then correct their
body orientation to align with their flight path.  For mosquitoes,
which fly without a strongly preferred body heading,
the heading might evolve passively over time as a result of active 
control of other aspects of the body
orientation, such as the roll.  
For instance, the body yaw may potentially be allowed to drift so long as it remains 
within a preferred range.
The mosquito's ability to generate thrust in arbitrary directions also means
that body yaw might be controlled because it dictates the orientation of the
insect's sensory system, not for its dynamic importance.
Identifying when the insect is actively maintaining its body heading is a potential
area for further investigation.  

We find little correspondence between body heading and the direction of body motion
or of body acceleration.  In addition, we find little correlation between body pitch angle and 
the insect's forward acceleration.  Although controlling
pitch angle is necessary for the insect to stay relatively level,
it does not dominate the generation of forward acceleration. 
It is likely that forward backward asymmetries in the
wing stroke directly generate such accelerations,
potentially with a paddling stroke [35].
%\cite{Ristroph2}.
The extent to which variation in pitch impacts the flight dynamics of the mosquito
is unclear.

While the exact pitch angle or body heading does not seem to impact the insect's motion,
the strong
correlation between sideways accelerations and roll-generated accelerations 
means that understanding the control of sideways force generation is closely related
to understanding how the mosquito generates roll torque and controls its body roll.
Roll stability, especially during steadily translating flight
with a large sideways component, is thus particularly relevant to 
mosquito dynamics. 
Questions of stability during flight are intertwined with the specifics of the
wing force generation model being analyzed.  When flight has a sideways velocity
component, the symmetry between the flow environments of the left and the right wing
is broken.  Quasi-steady models have not been experimentally validated under
steady sideways translation, and it is unclear what impact this translation has
on the sizes of the leading edge vortices that govern force generation.  Computational
simulations 
%\cite{Zhang} 
suggest that such translation generates changes in the
wing forces that are not captured by quasi-steady models [36].  
In addition, often flight stability
is studied for a model insect that is being perturbed from a hovering state or
from slow flight [37].
%(\cite{Hedrick}).
Even with effective aerodynamic models,
 analyses of flight stability about a hovering state may not
translate to flights with a steady nonzero sideways velocity component,
Since such flight appears to be ubiquitous in mosquitoes, 
this is an important stability question.
Better understanding control and stability of the roll degree of freedom,
as well as generalizing work on insect control beyond a hovering context to broader
flight regimes that include significant sideways flight components, will be important
to better understanding the exceptional flight abilities of mosquitoes, and 
of other small dipterans.  

\section{Acknowledgements}
We are indebted to Laura Harrington and Lauren Cator for providing mosquitoes and mosquito expertise.  
We also thank Itai Cohen and Tsevi Beatus
for use of the flight recording apparatus and for experimental support respectively.
Helpful discussions with John Guckenheimer, Ben Arthur and Derek Paley are also gratefully acknowledged.
This work was funded by the NSF under award number NSF CMS 0832782.

\end{document}